\documentclass[twocolumn,aps,nofootinbib]{revtex4} 

\usepackage{amsmath} 
\usepackage{graphicx}
\usepackage{dcolumn}
\usepackage{bm}

\newcommand{\nuc}[2]{$^{#1}${#2}} 

\begin{document} 

\title{Giant Resonances using Correlated Realistic Interactions: The Case for Second RPA}

\author{P.~Papakonstantinou} 
\email[Email:]{panagiota.papakonstantinou@physik.tu-darmstadt.de} 
\affiliation{Institut f\"ur Kernphysik, 
Technische Universit\"at Darmstadt, 
Schlossgartenstr.~9, 
D-64289 Darmstadt, Germany} 

\author{R.~Roth} 
\affiliation{Institut f\"ur Kernphysik, 
Technische Universit\"at Darmstadt, 
Schlossgartenstr.~9, 
D-64289 Darmstadt, Germany}

\begin{abstract} 
 
Lately we have been tackling the problem of describing nuclear collective excitations starting from correlated realistic nucleon-nucleon (NN) interactions. The latter are constructed within the Unitary Correlation Operator Method (UCOM), which explicitly considers short-range correlations in order to properly soften the short-range behaviour of realistic NN potentials. It has been concluded that first-order RPA with a two-body UCOM interaction (UCOM-RPA) is not capable, in general, of reproducing quantitatively the properties of giant resonances (GRs), due to missing higher-order configurations and long-range correlations as well as neglected three-body terms in the Hamiltonian. 
 
In the present paper we report results on GRs obtained by employing a UCOM interaction, based on the Argonne V18 potential, in Second RPA (SRPA). 
The same interaction is used to describe the Hartree-Fock (HF) ground state and the residual interactions. 
We find that the inclusion of second-order configurations -- which effectively dress the underlying HF single-particle states with self-energy insertions -- produces sizable corrections. 
The effect appears essential for a realistic description of GRs when using the UCOM. 
We argue that effects of higher than second order should be negligible. 
Therefore, the UCOM-SRPA emerges as a promising tool for consistent calculations of collective states in closed-shell nuclei. 
This is an interesting development, given that SRPA can accommodate more physics than RPA (e.g., fragmentation). 
Remaining discrepancies due to the missing three-body terms and self-consistency issues of the present SRPA model are pointed out.

\end{abstract} 

\maketitle


\section{Introduction}
 
Many-body approximations like Hartree-Fock (HF) and Random-Phase Approximation (RPA) (and their counterparts for open-shell nuclei, namely HF-Bogoliubov (HFB) and Quasiparticle-RPA (QRPA)) have allowed massive calculations of nuclear ground-state and excited-state properties throughout the nuclear chart. 
Such models are used in conjunction with effective nucleon-nucleon (NN) interactions.  
Indeed, the bare NN interaction induces strong correlations in the nuclear system -- most notably short-range correlations -- which cannot be described by simple model spaces such as those involved in HF, RPA, etc. 
One can find very good parameterizations of the effective NN force -- and there is now intense and coordinated activity towards the development of high-quality energy functionals to serve such a purpose -- but those have been phenomenological up to now, lacking a direct connection with the underlying bare interaction. 

The question remains as to whether it is possible to construct a global effective NN interaction starting from the bare one. There have been two recent attempts towards that direction. One is the construction of a low-momentum interaction, the so-called $V_{\mathrm{low-k}}$, by integrating out the high-momentum components of the bare one (thus softening its short-range behavior) using renormalization group techniques~\cite{BKS2003}. 
The other one 
is the Unitary Correlation Operator Method (UCOM)~\cite{FNR1998,NeF2003,RNH2004}, which deals explicitly with the short-range correlations and is described in the next Section. 
Applied to a realistic NN interaction, the UCOM 
produces a ``correlated" interaction, 
$\text{V}_{\text{UCOM}}$. 
Although constructed following different formalisms, $V_{\mathrm{low-k}}$ and $V_{\mathrm{UCOM}}$ have similar 
low-momentum matrix elements.%
\footnote{There remain important differences, however, as demonstrated, e.g., by the fact that the $V_{\mathrm{UCOM}}$ is able to produce stable (saturated) nuclear matter, while $V_{\mathrm{low-k}}$ is not.} 
Moreover, they do not depend strongly on the particular bare NN potential on which they are based, and to which 
they are phase-shift equivalent. 
The hope is to be able to employ such realistic but ``softened" potentials in many-body calculations. 

In the present work we focus on nuclear giant resonances (GRs) of closed-shell nuclei. 
First-order RPA with a two-body UCOM interaction has not been able to reproduce quantitatively the properties of all GRs%
~\cite{PPH2006,PRP2007}. 
Here we report results on GRs obtained using Second RPA (SRPA) and employing the correlated Argonne V18 interaction (UCOM-SRPA). 

It is not straightforward to perform SRPA calculations self-consistently -- in the sense that exactly the same interaction is used to describe the ground state and the residual couplings -- without conceptual problems. 
In typical SRPA applications in the past, phenomenological single-particle energies have been used and  G-matrix or phenomenological forces have been employed as residual interactions. (The real part of the SRPA self energy would then be discarded, since it would shift the already realistic single-particle energies.)  
Phenomenological density functionals, on the other hand,  
are typically fitted by using HF(B) and (Q)RPA results. 
Part of the long-range correlations 
affecting ground-state properties 
are then effectively taken into account  by the parameterization 
and higher-order effects are usually ignored. 
Employing such interactions in SRPA 
might result in overcounting of such effects.%
\footnote{One could, in principle, consider to fit their parameters using SRPA, but that would be a formidable task from a computational point of view. Note that computationally friendly zero-range interactions are not appropriate for large-scale SRPA.} 
Our correlated interaction, however, 
takes into account only short-range correlations; 
long-range correlations have to be described by extending  
the configuration space and one way to do that is SRPA. 

Let us note at this point that the RPA reaches its limits when confronted with problems such as the width and fine structure of GRs, the strength of some low-lying states, etc. 
The SRPA is a more appropriate theory to deal with such issues.

In the next section we outline the basic principles of the UCOM scheme. In Sec.~\ref{RPA} we review what we have learned so far by using the $V_{\mathrm{UCOM}}$
 in HF, perturbation theory, and first-order RPA calculations. In Sec.~\ref{srpa} we present the SRPA formalism and our new results. In Sec.~\ref{suma} we give a summary and perspectives. 

\section{The UCOM Hamiltonian} 
\label{UCOM} 

The basic idea of the UCOM is the explicit treatment of the interaction-induced 
short-range central and tensor correlations. These are imprinted into an 
uncorrelated many-body state $|\Psi\rangle$ (e.g., a Slater determinant) 
through a state-independent unitary transformation 
defined by the unitary correlation operator $C$, resulting in a correlated state 
$|\tilde{\Psi}\rangle = C |\Psi\rangle .  
$ 
The correlation operator $C$ is written as a product of unitary operators 
$C_{\Omega}$ and $C_{r}$ describing tensor and central correlations, respectively. 
Both are formulated as exponentials of a Hermitian generator, 
\begin{equation}
\label{eq:correlator}
  C = C_{\Omega} C_{r}
  = \exp [-\text{i} \sum_{i<j} g_{\Omega,ij} ]
    \exp [-\text{i} \sum_{i<j} g_{r,ij} ].
\end{equation}
The construction of the two-body generators $g_r$ and $g_{\Omega}$ follows the physical mechanisms 
by which the interaction induces central and tensor correlations. 
The short-range central correlations, caused by the repulsive core of the interaction, 
are introduced by a radial distance-dependent shift pushing nucleons apart from each other 
if they are within the range of the core. 
Tensor correlations between two nucleons are generated by a spatial shift perpendicular to the radial direction. 
For a given bare potential, 
the corresponding correlation functions are determined by an energy minimization in the two-body system for each 
$(S,T)$ channel. 

Matrix elements of an operator $O$ 
with correlated many-body states 
$|\tilde{\Psi}\rangle$  
can be equivalently written as matrix elements of a ``correlated" (transformed)  
operator $\tilde{O}=C^{\dagger}OC$ and uncorrelated many-body states 
$|\Psi\rangle$. 
Thus, one can work in simple Hilbert spaces (simple states) using correlated operators, 
rather than with bare operators and explicitly correlated states. 
By applying the 
transformation to a bare NN interaction, 
a phase-shift equivalent correlated interaction is obtained, is suitable for use in 
tractable model spaces~\cite{RNH2004,RHP2005,RPP2006}. 
The same transformation can then be applied to any 
other operator under study, as is needed for a consistent UCOM treatment. 

In an $A$-body system a correlated operator contains irreducible 
contributions to all particle numbers. 
The cluster expansion of a correlated operator reads 
\begin{equation} 
  \tilde{O} = C^{\dagger} O C = \tilde{O}^{[1]} + \tilde{O}^{[2]} + \cdots + \tilde{O}^{[A]}, 
\label{cexpansion}
\end{equation} 
where $\tilde{O}^{[n]}$ denotes the irreducible $n$-body contribution.  
In actual applications of the UCOM a two-body approximation is usually employed, 
i.e., three-body and higher-order terms of the expansion are neglected. 
Starting from the uncorrelated Hamiltonian $H$ for the $A$-body system,
consisting of the kinetic energy operator $T$ and a two-body potential $V$, 
the formalism of the UCOM is used to construct 
the correlated Hamiltonian in two-body approximation
\begin{equation}
  H^{C2} = {\tilde{T}}^{[1]} + {\tilde{T}}^{[2]} + {\tilde{V}}^{[2]}
  = T + V_{\text{UCOM}},
\end{equation}
where the one-body contribution comes only from the uncorrelated kinetic energy $\tilde{T}^{[1]}=T$. 
Two-body contributions arise from the correlated kinetic energy ${\tilde{T}}^{[2]}$ and the correlated 
potential ${\tilde{V}}^{[2]}$, which together constitute the phase-shift equivalent correlated interaction $V_{\text{UCOM}}$.

It has been verified that higher-order contributions due to short-range central correlations 
can be neglected in the description of nuclear structure properties \cite{RNH2004}. 
The tensor interaction, on the other hand, 
is long-ranged and thus generates long-range correlations 
in an isolated two-nucleon system. 
However, the long-range tensor 
correlations between two nucleons embedded in a many-nucleon system 
are suppressed by the presence of other nucleons, 
leading to a screening of the tensor correlations at large interparticle distances. 
In order to effectively describe the screening effect and at the same time justify the two-body approximation, 
the range of the tensor correlation function 
--- more precisely, 
the ``correlation volume" $I_{\vartheta}^{(S,T)}$~\cite{RHP2005} ---  
is restricted during the parameterization procedure. 
Restricting the range of the tensor correlator has another important function, 
namely to ensure that only 
state-independent, short-range correlations are described by the UCOM. 
By varying the correlation volumes --- the only parameters entering the formalism ---  
a family of correlators and respective correlated 
interactions are obtained. 

The question is then how to optimize these parameters in order to 
best describe the screening effect and the separation of the two types 
of correlations. 
As demonstrated in Ref.~\cite{RHP2005}, 
this can be done 
with the help of exact few-body calculations. 
In particular, 
the values can be chosen so as to best describe the binding energies of 
\nuc{3}{H} and \nuc{4}{He} 
within the no-core shell model 
($I_{\vartheta}^{(1,0)}=0.09$~fm$^3$ 
for the Argonne V18 potential). 
For such a choice of tensor correlator range the 
missing genuine three-nucleon interaction and  the omitted higher-order 
terms of  the cluster expansion of the correlated Hamiltonian effectively cancel 
each other. 
As was subsequently shown within many-body perturbation theory~\cite{RPP2006}, 
and verified by RPA calculations~\cite{BPR2006}, this cancelation remains at work 
throughout the nuclear chart, as far as the binding energy is concerned 
(see also next Section).%
\footnote{The $V_{\mathrm{UCOM}}$ has a strong momentum dependence, even when it is based on a local potential like Argonne V18. That is why it can perform reasonably well without an additional three-body term.} 

In this work we will use the correlated Argonne V18 potential with 
$I_{\vartheta}^{(1,0)}=0.09$~fm$^3$. 
No tensor correlator is employed in the triplet-odd channel, 
where the tensor interaction is much weaker. 
We start from a Hamiltonian which consists of the intrinsic 
kinetic energy $T_{\text{int}}$ and the $V_{\text{UCOM}}$ interaction 
derived from the Argonne V18 potential including the Coulomb potential, 
\begin{equation} 
\label{eq:hintr}
  {H}_{\text{int}} 
  = T - T_{\text{cm}} + V_{\text{UCOM}} 
  = T_{\text{int}} + V_{\text{UCOM}} \;,
\end{equation} 
in two-body approximation. 
It is the two-body Hamiltonian $H_{\text{int}}$ 
that has been used in Hartree-Fock (HF), perturbation-theory, and RPA calculations in Refs.~\cite{RPP2006,PPH2006,PRP2007} 
and that will be employed in this work too. 
In practice, two-body matrix elements in a harmonic-oscillator basis are the input 
to such calculations.  

\section{Applications in spherical nuclei: recent lessons} 
\label{RPA} 

Using the $V_{\mathrm{UCOM}}$ in HF calculations we obtained bound nuclei throughout the nuclear chart~\cite{RPP2006}. 
The tensor correlations play an important role in this. 
Note, though, that using the UCOM we aim to treat explicitly only the state-independent 
short-range correlations; long-range correlations should be 
described by the model space. 
This tells us already 
that the UCOM-based HF is not enough, since a Slater-determinant 
wavefunction is unable to describe correlations. It is found indeed 
that 
 the binding energies are underestimated 
by about 4~MeV per nucleon. The charge radii are underestimated too. 
While the Fermi energy is correctly reproduced, 
the level spacing of the 
single-particle states is too small. 

Second-order perturbation theory constitutes 
a tractable extension to the ``zero-order" description provided by HF and was employed in Ref.~\cite{RPP2006}. 
The very good description of nuclear binding energies 
achieved within perturbation theory for nuclei from $^4$He to $^{208}$Pb shows that 
the cancellation between the omitted three-body terms 
of the cluster expansion and genuine three-body correlations and terms of the interaction 
works throughout the nuclear chart as far as the binding energies are concerned. 
Charge radii 
are still underestimated within perturbation theory, suggesting that the above-mentioned cancellation 
does not work for all observables and that 
supplementing our two-body Hamiltonian with a three-body term to take account of 
missing effects may be necessary for realistic nuclear-structure calculations.  
Higher than second-order corrections are found to be small. 

The $V_{\text{UCOM}}$ has also been employed in 
standard, self-consistent RPA calculations 
to study nuclear giant resonances~\cite{PPH2006}. 
The ground state was described by the 
uncorrelated HF state, as usual. 
The isoscalar (IS) giant monopole resonance (GMR), 
the isovector (IV) giant dipole resonance (GDR), 
and the IS giant quadrupole resonance (GQR) were examined. 
Highly collective states were obtained for various closed-shell nuclei ranging 
from \nuc{16}{O} to \nuc{208}{Pb}. A reasonable agreement with the 
experimental centroid energies of the IS GMR was achieved. By contrast, 
the energies of the IV GDR and the IS GQR 
were overestimated by several MeV.  

Obviously, the $V_{\text{UCOM}}$ is not a traditional effective interaction. 
Partly because no long-range correlations are (effectively) included in the UCOM, 
the corresponding nucleon effective mass in nuclear matter 
obtained in a HF calculation is very low 
(around half the bare nucleon mass). 
This is confirmed by the HF results in finite nuclei, in particular the small level density. 
It is also manifested by the above-mentioned RPA results 
on the GQR and GDR centroids. 
It follows then that, besides the possible important 
role of missing three-body terms in the Hamiltonian, another source of our 
failure to describe nuclear collective states quantitatively with UCOM-RPA can be  
residual long-range correlations. 

The standard RPA is based on the assumption that the 
true RPA ground state can be approximated by the HF ground state. 
It is not obvious that this assumption holds when the $V_{\text{UCOM}}$  
is used, given the large correction to the HF binding energies 
due to second-order~\cite{RPP2006} and RPA~\cite{BPR2006} correlations. 
Therefore, in Ref.~\cite{PRP2007} 
the effect of explicit RPA ground-state correlations 
on the results for GRs was examined. 
To this end, 
a renormalized RPA version was used~\cite{CPS1996,CGP1998,VKC2000}. 
The effect on the properties of 
GRs was found to be rather small. 
It is concluded that first-order RPA with a two-body UCOM Hamiltonian cannot describe quantitativly 
the properties of GRs. 

Up to now we have assumed that residual three-body 
forces can be neglected, based upon the fact that they contribute only marginally  
to the ground-state energy as calculated within many-body perturbation theory~\cite{RHP2005,BPR2006}. 
This is not necessarily a valid assumption. 
A simple phenomenological zero-range three-body force can be constructed in order to be used along with 
the correlated two-nucleon interaction in future calculations. 
Preliminary results show that 
by using such a three-body force it is possible to improve on the description of observables such 
as nuclear radii and resonance energies while retaining the good reproduction of the experimental binding energies. 

Another important issue with RPA is that only one-particle-one-hole excitations are taken into account 
and the coupling to higher-order configurations 
($2p2h$ and beyond) is neglected. 
One can include higher-order configurations, starting with two-particle-two-hole 
within SRPA. Given that an extended model space is of great importance 
when using the $V_{\text{UCOM}}$, it is imperative to examine the effect. 

 
\section{Second RPA} 
\label{srpa} 

\subsection{Formalism} 

We will use the SRPA as it was formulated in Ref.~\cite{Yan1987} in analogy to RPA.  
Excited states $|\nu\rangle $ of energy $E_{\nu}=\hbar\omega_{\nu}$ with respect to the ground state $|0\rangle$  
\begin{equation} 
|\nu\rangle = Q_{\nu}^{\dagger} |0\rangle , 
\end{equation} 
are considered as combinations of $1p1h$ and $2p2h$ configurations. 
(We omit angular momentum coupling to keep the notation simple.) 
The corresponding creation operators are then written as 
\begin{eqnarray}  
Q_{\nu}^{\dagger} &=& 
\mbox{$\sum_{ph}$} X_{ph}^{\nu} O^{\dagger}_{ph} 
- \mbox{$\sum_{ph}$} Y_{ph}^{\nu} O_{ph} 
\nonumber \\ 
&&  
 + \mbox{$\sum_{p_1h_1p_2h_2}$} \mathcal{X}_{p_1h_1p_2h_2}^{\nu} O^{\dagger}_{p_1h_1p_2h_2} 
\nonumber \\ 
&&  
 - \mbox{$\sum_{p_1h_1p_2h_2}$} \mathcal{Y}_{p_1h_1p_2h_2}^{\nu} O_{p_1h_1p_2h_2} 
, 
\end{eqnarray} 
where $O^{\dagger}_{ph}$ 
creates a $ph$ state and 
$O^{\dagger}_{php'h'}$ 
creates a $2p2h$ state. 
The SRPA ground state, which is the vacuum of the annihilation operators $Q_{\nu}$, 
is approximated by the HF ground state. 
The forward ($X$, $\mathcal{X}$) 
and backward ($Y$, $\mathcal{Y}$) 
amplitudes are then given by the SRPA equations  
in $ph\oplus 2p2h-$space 
\begin {equation} 
\left( \begin{array}{cc|cc}  
A                & \mathcal{A}_{12} & B  & 0 \\ 
\mathcal{A}_{21} & \mathcal{A}_{22} & 0  & 0 \\ \hline  
-B^{\ast}        &  0               & -A^{\ast} & -\mathcal{A}^{\ast}_{12} \\ 
   0             &  0               & -\mathcal{A}_{21}^{\ast} & -\mathcal{A}^{\ast}_{22} \\ 
\end{array} 
\right) 
\left( 
\begin{array}{c} 
X^{\nu} 
\\ 
\mathcal{X}^{\nu}  
\\ 
\hline  
Y^{\nu} 
\\ 
\mathcal{Y}^{\nu}  
\end{array} 
\right) 
= \hbar\omega_{\nu}  
\left( 
\begin{array}{c} 
X^{\nu} 
\\ 
\mathcal{X}^{\nu}  
\\ 
\hline  
Y^{\nu} 
\\ 
\mathcal{Y}^{\nu}  
\end{array} 
\right) 
\label{Esrpa}  
, \end{equation} 
where $A$ and $B$  
are the usual RPA matrices, 
$\mathcal{A}_{12}$ describes the coupling between $ph$ and $2p2h$ states 
and 
$\mathcal{A}_{22}$ contains the $2p2h$ states and their interactions. 
If we neglect the coupling amongst those states, $\mathcal{A}_{22}$ is diagonal and its elements are equal to the 
unperturbed $2p2h$ energies, 
\begin{equation} 
\mathcal{A}_{22} = \delta_{p_1p_1'}\delta_{h_1h_1'}\delta_{p_1p_1'}\delta_{h_1h_1'}(e_{p_1}+e_{p_2}-e_{h_1}-e_{h_2}) 
\label{Ea22} 
\end{equation} 
($e_i$ are the HF single-particle energies). 

The dimension $N$ of the SRPA matrix, Eq.~(\ref{Esrpa}), can be very large. 
Fortunately, the SRPA matrix is also sparse. When the approximation (\ref{Ea22}) is employed, most of its elements are zero. 
Thus it is possible to store all its finite matrix elements and then use a Lanczos procedure to obtain only the 
eigenvectors of interest. 

The SRPA problem, Eq.~(\ref{Esrpa}), can be reduced to an energy-dependent eigenvalue problem of the dimension of the RPA matrix (see, e.g., Ref.~\cite{Wam1988}). Therefore, it can be viewed as an RPA problem with an energy-dependent interaction. In general, the reduction procedure involves the inversion of a complex matrix in the large $2p2h$ space, but when $\mathcal{A}_{22}$ is diagonal, that is reduced to a trivial complex-number inversion. There are ways to solve an energy-dependent eigenvalue problem~\cite{BAD1990,All1993}.  
An efficient alternative is to employ the response-function formalism. Then, instead of explicitly solving the eigenvalue problem, one can obtain directly the strength distribution of interest \cite{Wam1988,All1993}. We have used this technique as well.  

It has been shown formally that the total strenght $m_0$ and the first moment of the strength distribution $m_1$ are the same in the present SRPA as in RPA~\cite{AdL1988}. 
However, when based on the HF ground state, the SRPA is not fully self-consistent and symmetry-conserving, contrary to the RPA based on the HF ground state. 
It has been pointed out~\cite{TSA1988} that it misses a class of second-order effects, related to ground state correlations. 
The missing effects may be important, especially for the less collective low-lying states.  
In principle, it is possible to combine the SRPA with a correlated ground state~\cite{TSA1988,DNS1990,GGC2006} 
for a most complete theoretical treatment of nuclear excitations, but that is beyond the purposes of the present work.  

\subsection{Results} 

We have used the correlated Argonne V18 interaction and a single-particle basis of 11 oscillator shells and we have examined the IS monopole (ISM), IV dipole (IVD) and IS quadrupole (ISQ) response of the nuclei $^{16}$O and $^{40}$Ca. 
The convergence of the GR sum rules $m_0$ and $m_1$ and centroids is rather good for the present basis (within about 1~MeV for the centroids). 
The total number of eigenvalues is $10^{4-5}$ for the cases presented here (it can be larger e.g. for heavier nuclei), but  
less than 300 eigenstates are sufficient to describe the region of the GRs. 
We use standard single-particle transition operators~\cite{PPH2006}. 
We present our results in comparison with experimental data. 
The experimental centroids $m_1/m_0$ of the IS GMR and the IS GQR were taken from 
Refs. \cite{LCY2001} (\nuc{16}{O}) and \cite{YLC2001} (\nuc{40}{Ca}). 
Photoabsorption cross sections were found in 
Refs. \cite{Ahr1975,LNH1987} (\nuc{16}{O}) and \cite{Vey1974} (\nuc{40}{Ca})%
\footnote{Data are available from the CDFE database, 
http://cdfe.sinp.msu.ru/services/gdrsearch.html} 
and strength distributions and centroids of 
the IV GDR were evaluated from those. 

We have verified that in SRPA the total $m_0$ is almost the same as in RPA. The total $m_1$ is smaller by more than 20\%. This is probably because we do not calculate the full spectrum. A non-negligible part of the total $m_1$ may be distributed among the large number of weak excitations lying at high energies. 
\begin{figure*}[t]\centering
\includegraphics[angle=-90,width=120mm]{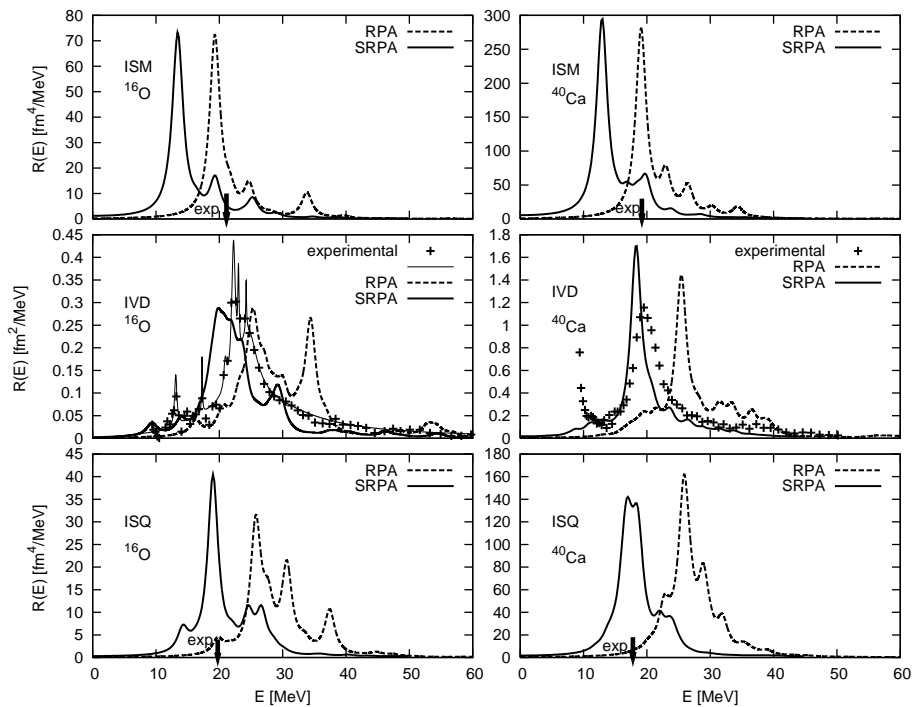} 
\caption{The IS monopole (top), IV dipole (middle) and IS quadrupole (bottom) strength distributions for the nuclei $^{16}$O (left) and $^{40}$Ca (right) within RPA and SRPA, compared with experiment (for references see text). 
The single-particle basis consists of 11 oscillator shells.  
The calculated distributions (RPA and SRPA) have been folded with a Lorenzian with a width of 2~MeV and thus the SRPA fragmentation is not visible. 
\label{Fall}}
\end{figure*}

It has been shown~\cite{ToS2004} that the spurious state related to the CM motion will generally not be exactly seperated from the physical spectrum, when SRPA is based on the HF ground state. In order to quantify this problem, we have examined the behaviour of the IS dipole response. We found that the spurious state appears at about 5~MeV. We used a transition operator of the usual radial form ($\propto r^3 - \frac{5}{3}\langle r^2 \rangle r$) and its uncorrected form ($\propto r^3$) and found that the spectrum beyond the spurious state is practically the same and can be considered uncontaminated. 

In Fig.~\ref{Fall} we show the ISM, IVD and ISQ strength distributions for the two nuclei. 
Note that, for presentation purposes, the calculated distributions (RPA and SRPA) have been folded with a Lorenzian with a width of 2~MeV. Thus, all peaks have acquired an artificial width (which for some low-lying dipole states may be too large) and the SRPA fragmentation is not visible. 
In all cases, the SRPA centroid energies are much lower than the RPA ones. 
The reason for the difference between the RPA and SRPA results -- even for such collective $ph$ excitations like the GRs considered here -- is to a large extent that, within SRPA, the coupling of single-particle states with virtual phonons is implicitly taken into account.  
The inclusion of second-order configurations within SRPA effectively dresses the underlying HF single-particle states with self-energy insertions and brings them closer to each other energetically, thereby lowering the underlying $ph$ energies. It is an important physical effect which cannot be ignored when using completely ``undressed" (with respect to long-range correlations) HF states like the ones produced by the $V_{\mathrm{UCOM}}$. 
In this scheme the undressed HF energies are viewed as auxiliary model quantities which should not be directly compared with experiment. 
  
Let us look at the results in more detail. 
In the middle panels of Fig.~\ref{Fall} the RPA and SRPA strength distributions are shown for the IV GDR, along with those extracted from experimental data (there has been no ad hoc renormalization imposed). 
We observe that the IV GDR is more realistically reproduced within SRPA than within RPA. 
Its centroid energy is somewhat underestimated. 
In the lower panels we show the ISQ strength distributions. The RPA and SRPA results are shown and the experimental centroids of the IS GQR are indicated. The agreement of the SRPA results with experiment is very good. 
It appears as though, once coupling to higher-order configurations is taken into account, a realistic effective mass is restored. 
 
In the upper panels of Fig.~\ref{Fall} we show the ISM strength distributions. The energies of the IS GMR are underestimated within SRPA. This is another indication that there are missing three-body effects and our two-body interaction should be supplemented with a three-body term to describe them. 
Normally, residual three-body corrections should affect the IS GMR most of all, since it is a compression mode. They should affect less strongly the IV GDR, where the nuclear interior plays a lesser role, and less the IS GQR, which is a surface mode. These physical arguments could serve as a guide for the construction of an appropriate effective three-body term. 

In the above, the approximation (\ref{Ea22}) has been used. It has been verified that inclusion of the couplings amongst the $2p2h$ states produces negligible corrections. Note that those couplings constitute higher-order effects. The indications that we had from our perturbation-theory results, that corrections beyond second order are small, are thus confirmed.

\section{Summary and perspectives} 
\label{suma} 

We have employed a correlated interaction $V_{\text{UCOM}}$ based on the Argonne V18 potential in SRPA calculations of nuclear GRs. 
Short-range correlations are explicitly taken into account.  
The same interaction is used to describe the Hartree-Fock (HF) ground state and the residual interactions. 
We found that the second-order configurations 
produce sizable corrections with respect to first-order RPA. 
They do so by effectively dressing the underlying HF single-particle states with self-energy insertions. 
The effect appears essential for a realistic description of GRs when using the $V_{\text{UCOM}}$. 
Effects of higher than second order should be negligible. 
Therefore, the UCOM-SRPA model emerges as a promising tool for consistent calculations of collective states in closed-shell nuclei. 
This development is interesting, given that SRPA can accommodate more physics than RPA (e.g., fragmentation width and fine structure of GRs). 
Remaining discrepancies, regarding in particular the IS GMR, can be attributed to missing three-body effects. 
Self-consistency issues of the present SRPA formulation were also pointed out. 

Up to now we have considered mostly the centroids of GRs, but their decay properties can also be studied within UCOM-SRPA. 
Heavier nuclei and  
low-lying states will be a topic for future work as well. 

\section*{Acknowledgments}

Work supported by the Deutsche Forschungsgemeinschaft, 
contract SFB 634.


\end{document}